\newcommand{\be}{\begin{equation}}\newcommand{\ee}{\end{equation}}
\newcommand{\bea}{\begin{eqnarray}}\newcommand{\eea}{\end{eqnarray}}
\newcommand{\nn}{\nonumber}\newcommand{\p}[1]{(\ref{#1})}
\newcommand{\lb}{\label}

\newcommand\s{\scriptscriptstyle}
\newcommand{\ab}{{\alpha\beta}}
\newcommand{\pab}{\partial_\ab}
\newcommand{\ta}{\theta^\alpha}
\newcommand{\tb}{\theta^\beta}
\newcommand{\bta}{\bar{\theta}^\alpha}
\newcommand{\btb}{\bar{\theta}^\beta}
\newcommand{\tao}{\theta^\alpha_1}
\newcommand{\tbo}{\theta^\beta_1}
\newcommand{\tat}{\theta^\alpha_2}
\newcommand{\ts}{(\theta)^2}
\newcommand{\bts}{(\bar{\theta})^2}
\newcommand{\tbt}{(\theta\bar{\theta})}
\newcommand{\too}{(\theta_1\theta_1)}
\newcommand{\ttt}{(\theta_2\theta_2)}
\newcommand\cD{{\cal D}}
\newcommand\bcD{\bar{\cal D}}
\newcommand{\cDa}{\cD_\alpha}
\newcommand{\cDb}{\cD_\beta}
\newcommand{\bcDa}{\bcD_\alpha}
\newcommand{\bcDb}{\bcD_\beta}
\newcommand\cDao{{\cal D}_\alpha^1}
\newcommand\cDat{{\cal D}_\alpha^2}

\newcommand\cQa{{\cal Q}_\alpha}
\newcommand\bcQa{\bar{\cal Q}_\alpha}

\newcommand{\Da}{D_\alpha}

\newcommand{\bDa}{\bar{D}_\alpha}

\newcommand{\Ds}{(D)^2}
\newcommand{\bDs}{(\bar{D})^2}
\newcommand{\DbD}{(D\bar{D})}
\newcommand{\Dao}{D_\alpha^1}
\newcommand{\Dat}{D_\alpha^2}
\newcommand{\Dbo}{D_\beta^1}

\newcommand{\Doo}{(D^1D^1)}
\newcommand{\Dtt}{(D^2D^2)}

\documentstyle[12pt]{article}

\begin{document}
\renewcommand{\thefootnote}{\fnsymbol{footnote}}
\hfill hep-th/9905108\\
\vspace{0.5cm}

\begin{center}
{\large\bf
PARTIAL SPONTANEOUS BREAKDOWN OF 3-DIMENSIONAL N=2 SUPERSYMMETRY}

\vspace{0.5cm}
{\bf  B.M. Zupnik}\footnote{On leave on absence from the Institute
of Applied Physics, Tashkent State University, Uzbekistan}\\
{\it Bogoliubov Laboratory of Theoretical Physics, Joint Institute
for Nuclear Research, 141980, Dubna, Russia}\\
e-mail: zupnik@thsun1.jinr.ru\\
\end{center}

\begin{abstract}
The superfield models with the partial spontaneous breaking of the global 
$D{=}3$, $~N{=}2$ supersymmetry are discussed. The abelian gauge model 
describes  low-energy interactions of the real scalar field with the 
$3D$ vector and fermion fields. We introduce the new Goldstone-Maxwell 
representation of the $3D$ gauge superfield and show that the partial 
spontaneous breaking  $N{=}2\rightarrow N{=}1$ is possible for the 
non-minimal self-interaction of this modified gauge superfield including 
the linear Fayet-Iliopoulos term. The dual description of the partial 
breaking in the model of the self-interacting Goldstone chiral superfield 
is also considered. These models have the constant vacuum solutions and 
describe, respectively, the interactions of the $N{=}1$ Goldstone 
multiplets of the $D2$-brane or supermembrane with the additional
massive multiplets.
\end{abstract}



\renewcommand{\thefootnote}{\arabic{footnote}}
\setcounter{footnote}0
\setcounter{equation}0
\section{Introduction}
Standard mechanisms of spontaneous breaking of the global $D{=}4,~
N{=}1$ supersymmetry ($SBGS$) are connected with the constant
vacuum solutions for the auxiliary components of chiral and gauge
superfields (see reviews \cite{WB}-\cite{GGRS}). The constant 
$SBGS$ solutions are possible in the very restrictive class of the
self-interacting models of chiral superfields.
In particular, $SBGS$ is not possible for the non-trivial self-interaction
of the single chiral superfield. The Fayet-Iliopoulos $(FI)$ mechanism
consists in adding the linear term to the action of the
$N{=}1$ abelian gauge theory, however, this term does not guarantee
automatically the appearance of the $SBGS$-solution for any gauge-matter
interaction. These standard mechanisms are not very
flexible, so the search of new approaches to this problem is desirable,
especially for the extended supersymmetry or supersymmetries in low
dimensions which have some specific features. The problems of the 
spontaneous breaking of local supersymmetries will not be discussed
in this paper.

The standard linear supermultiplets (standard superfields) are not
convenient for the description of the  partial spontaneous  breaking
 of the exended global supersymmetries $(PSBGS)$ when the invariance
with respect to the part of supercharges remains  unbroken. In particular, 
the constant solutions with a degenerate structure of the auxiliary fields
are forbidden  in many cases. The Goldstone-fermion models with the 
partial spontaneous breaking of the  $D{=}4,~N{=}2$ \cite{HLP} or
$D{=}3,~N{=}2$ \cite{AGIT} supersymmetries  have been constructed using 
the topologically non-trivial classical solutions preserving the one half
of supercharges. These models have been also studied in the method of 
nonlinear realizations of supersymmetries \cite{VA}-\cite{IK}
using  superfields of the unbroken supersymmetry.

Recently the abelian gauge model with two $FI$-terms have been used 
to break spontaneously $D{=}4,~N{=}2$ supersymmetry to its $N{=}1$
subgroup \cite{APT}-\cite{IZ}. This model describes the non-minimal 
interactions of the complex scalar field with the fermion
and $U(1)$-gauge fields. In the  $D{=}4,~N{=}2$ superspace these
interactions correspond to the holomorphic action of the Goldstone-Maxwell
chiral superfield $W$ satisfying the modified superfield 2-nd order
constraints. In comparison to the original constraints of the $N{=}2$ vector 
multiplet \cite{GSW}, these constraints contain the constant terms which
guarantee the appearance of the unusual constant imaginary part of the 
isovector auxiliary component and the  Goldstone fermion component in the
the Goldstone-Maxwell superfield.

The more early example of the Goldstone-type constraint has been
considered in the model with the partial breaking of the $D{=}1,~N{=}4$
supersymmetry \cite{IKP}. Thus, these constraints introduce a new type
of the supersymmetry representations with the linear Goldstone $(LG)$ 
fermions. In distinction with the Goldstone fermions of the nonlinear 
realizations which transforms linearly only in the unbroken supersymmetry,
the $LG$-fermions have their  partners in the  supermultiplets of the whole 
supersymmetry. The nonlinear deformation of the standard constraints is also
possible \cite{GGRS}, however, we shall discuss only constant terms
in the modified constraints which are connected with the spontaneous
breaking of supersymmetries. It will be shown that the models with the $LG$ 
vector multiplet and the corresponding dual scalar multiplet solve the 
problem of the partial spontaneous breaking of the $D{=}3,~N{=}2$ 
supersymmetry. Recently these problems have been considered in the 
framework of the $N{=}1$ superspace \cite{IK}.

The coordinates of the full $D{=}3,~N{=}2$ superspace are
\be
z=(x^\ab,\ta ,\bta )~,\lb{A1}
\ee
where $\alpha,\beta$ are the spinor indices of the group $SL(2,R)$.
The spinor representation of the coordinate is connected with the
vector representation via the $3D$ $\gamma$-matrices $x^\ab{=}(1/2)x^m
(\gamma_m)^\ab$. The algebra of spinor derivatives in this superspace has
the following form:
\bea
&& \{\cDa ,\bcDb\}=i\pab+i
\varepsilon_\ab Z~,\lb{A2}\\
&&  \{\cDa,\cDb\}=0~,\qquad \{\bcDa,\bcDb\}=0~,
\lb{A3}
\eea
where $Z$ is the real central charge and
\bea
&&\cDa=\Da + {i\over2}\bar{\theta}_\alpha Z~,\qquad \Da =\partial_\alpha +
{i\over2}\btb\pab~,\nn\\
&&\bcDa=\bDa -{i\over2}\theta_\alpha Z~,\qquad \bDa =\bar{\partial}_\alpha
+{i\over2}\tb\pab.\lb{A4}
\eea
We shall use mainly the spinor derivatives without the central charge
$\Da$ and $\bDa$.

The corresponding generators of the $N{=}2$ supersymmetry are
\be
\cQa =Q_\alpha +{1\over2}\bar{\theta}_\alpha
Z~,\qquad \bcQa=\bar{Q}_\alpha -{1\over2}\theta_\alpha Z~.
\lb{A4b}
\ee

The $N{=}2$ supersymmetry algebra  is covariant with respect to the
$U_R(1)$ transformations
\be
\theta^\alpha~\rightarrow~e^{i\rho}\theta^\alpha~,\qquad
\bta~\rightarrow~e^{-i\rho}\bta~.\lb{A4c}
\ee

 We shall consider the following
notation for the bilinear combinations of spinor coordinates and
differential operators:
\bea
&&\ts={1\over2}\theta_\alpha \theta^\alpha~,\quad \bts =
{1\over2} \bar{\theta}^\alpha \bar{\theta}_\alpha~,
\lb{A5}\\
&&\tbt ={1\over2}\ta\bar{\theta}_\alpha~,\qquad\Theta^{\alpha\beta}
={1\over2}  [\theta^\alpha\bar{\theta}^\beta +\alpha  \leftrightarrow
\beta ]~,\lb{A6}\\
&& \Ds ={1\over2}D^\alpha D_\alpha~,\qquad \bDs ={1\over2}
   \bar{D}_\alpha \bar{D}^\alpha~,\lb{A7b}\\
   &&\DbD ={1\over2}D^\alpha\bar{D}_\alpha~, \qquad D_{\alpha\beta}=
{1\over2}([D_\alpha,\bar{D}_\beta]+ \alpha  \leftrightarrow\beta)~.
\lb{A7}
\eea
and the useful relations
\bea
&& D_\alpha\bar{D}_\beta={i\over2}\pab +
\varepsilon_\ab(D\bar{D})+{1\over2} D_\ab~,\lb{A7c}\\
&& \DbD^2={1\over8}\partial^\ab(\pab-iD_\ab)+{1\over2}\Ds\bDs~,\lb{A8b}\\
&&D_\ab\Ds=i\pab \Ds~,\lb{A8c}\\
&& (D)^2 (\theta)^2=1~,\qquad  (\bar{D})^2 (\bar{\theta})^2=1~,\qquad
\DbD \tbt=-{1\over2}~. \lb{A8}
\eea

The integration measures in the full and chiral superspaces are
\be
d^7z=d^3x(D)^2(\bar{D})^2~,\qquad d^5\zeta =d^3x_L(D)^2~.\lb{A9}
\ee
They have $R$-charges 0 and $-2$, respectively. The complex chiral
coordinates can be constructed by the analogy with $D{=}4$
\be
\zeta=(x_L^\ab,\ta)~,\qquad x_L^\ab=x^\ab+i\Theta^\ab~.\lb{A9b}
\ee

It is convenient to use the following rules of conjugation for any
operators
\cite{GGRS}:
\be
(X Y)^\dagger=Y^\dagger X^\dagger~,\qquad [X,Y\}^\dagger=-(-1)^{p(X)p(Y)}
[X^\dagger,Y^\dagger\}~,\lb{conj1}
\ee
where $[X,Y\}$ is the graded commutator and $p(X)=\pm1$ is the
$Z_2$-parity. The action of the differential operator $X$ on some function
$f(z)$ and the corresponding conjugation are defined as follows:
\be
X f\equiv [X,f\}~\Rightarrow~(Xf)^\dagger=-(-1)^{p(X)p(f)}X^\dagger
f^\dagger~.\lb{conj2}
\ee
(Remark that the alternative convention of conjugation 
$(Xf)^\dagger=(-1)^{p(X)p(f)}X^\dagger f^\dagger$ is also possible.)

Consider the conjugation rules for the spinor coordinates and derivatives
\bea
&& (\ta)^\dagger=\bta~,\quad [\ts]^\dagger=\bts~,\quad \tbt^\dagger=-\tbt
~,\lb{conj3}\\
&&\Da^\dagger=\bDa~,\quad [\Ds]^\dagger=\bDs~,\quad\DbD^\dagger=-\DbD~.
\lb{conj4}
\eea

It is possible to introduce the real $N{=}2$ spinor coordinates
$\theta^\alpha_i=(\theta^\alpha_i)^\dagger $
\bea
&&\theta^\alpha ={1\over\sqrt{2}}(\theta^\alpha_1 +i\theta^\alpha_2)~,
\qquad\bar{\theta}^\alpha ={1\over\sqrt{2}}(\theta^\alpha_1 -
i\theta^\alpha_2)~,\lb{A11}\\
&&  (\theta)^2={1\over2}[(\theta_2 \theta_2)-
(\theta_1 \theta_1) -2i (\theta_1\theta_2)]~,\qquad (\theta_i\theta_k)
\equiv {1\over2}\theta^\alpha_i\theta_{k\alpha}=(\theta_k\theta_i)~,
\lb{A11b}\\
&& (\theta\bar{\theta})={1\over2}[ (\theta_1\theta_1)+
(\theta_2\theta_2)]~,\qquad (\theta_i\theta_k)^\dagger=-(\theta_i\theta_k)
\lb{A11c}~,\\
&&\Theta^\ab={i\over2}(\tat\tbo +\alpha\leftrightarrow\beta)~,\qquad
(\Theta^\ab\lb{A11d})^\dagger=\Theta^\ab
\eea
and the corresponding real spinor derivatives
\bea
&&\cDao=\Dao +{1\over2}\theta_{2\alpha}Z~,\qquad \cDat=\Dat -{1\over2}
\theta_{1\alpha}Z~,\lb{A12}\\
&& D^1_\alpha={1\over\sqrt{2}}(D_\alpha +\bar{D}_\alpha)~,\qquad
D^2_\alpha={i\over\sqrt{2}}(D_\alpha -\bar{D}_\alpha)~,\lb{A13}\\
&& \{D^1_\alpha,D^1_\beta\}=\{D^2_\alpha,D^2_\beta\}=i\partial_{\alpha
\beta}~,\qquad
\{D^1_\alpha,D^2_\beta\}=0~.\lb{A14}
\eea

The  $D{=}3,~ N{=}2$ gauge theories have been considered, for instance, in 
refs.\cite{Si}-\cite{BHO}. The non-minimal self-interaction of the $U(1)$ 
gauge supermultiplet in this case is equivalent to the interaction of the 
$3D$ linear multiplet. We shall analyse the  modified $LGM$-constraints 
for the $3D$ gauge multiplet. The corresponding real $3D$ superfield
describes the scalar field interacting with the Goldstone fermions and the
vector field.

In sect.\ref{D} we discuss the prepotential solution for the $LGM$ 
supermultiplet which contains additional terms manifestly depending 
on the spinor coordinates and some complex constants  playing the role 
of moduli in the vacuum state of the theory together with the constant of 
the $FI$-term. Using this representation in the non-minimal gauge action 
one can obtain the constant vacuum solutions with the partial  spontaneous 
breaking of the $D{=}3,~N{=}2$ supersymmetry. Note that the 
supersymmetry algebra  is modified on the $LGM$ prepotential $V$ by analogy
with the similar modified transformations of the $4D$ gauge fields or 
prepotentials in refs.\cite{FGP,IZ}.

The sect.\ref{E} is devoted to the description of  $PSBGS$ in the
 interaction of the $LG$-chiral superfield which is dual to the 
 interaction of the $LGM$  superfield. This manifestly supersymmetric
 action depends on the sum of the chiral and antichiral superfields and
some constant term bilinear in the spinor coordinates. The non-usual
transformation of the basic $LG$-chiral superfield satisfies the
supersymmetry algebra with the central-charge term.

The $N{=}1$ supermembrane and $D2$-brane actions \cite{IK} can be analysed
in our approach using the decompositions of $N{=}2$ superfields in the 2-nd 
spinor coordinate $\theta^\alpha_2$. In sect.\ref{F}, we consider the 
$N{=}1$ components of the extended superfields and the covariant conditions 
which allow us to express the additional degrees of freedom in terms of the 
Goldstone superfields.

\setcounter{equation}0
\section{\lb{B}Vector multiplet in $D{=}3,~ N{=}2$ supersymmetry}

The  $D{=}3,~ N{=}2$ gauge theory \cite{Si,ZP1,AHISS,BHO} is analogous to 
the well-known $D{=}4,~ N{=}1$ gauge theory, although the three-dimensional
case has some interesting peculiarities which are connected with the
existence of the topological mass term and duality between the $3D$-vector 
and chiral multiplets. We shall consider the basic superspace with
$Z{=}0$.

The abelian $U(1)$-gauge prepotential $V(z)$  possesses the gauge
transformation
\be
\delta V=\Lambda +\bar{\Lambda}~,\lb{B5}
\ee
where the chiral and anti-chiral parameters are considered
\be
\bar{D}_\alpha\Lambda=0~,\qquad      D_\alpha\bar{\Lambda}=0~.\lb{B6}
\ee

The $D{=}3,~ N{=}2$ vector multiplet is described by the real linear
superfield 
\be
W(V)= i(D\bar{D})V\lb{B7}
\ee
satisfying the basic constraints
\be
(D)^2W=(\bar{D})^2W=0~.\lb{B8}
\ee

The additional useful relations for this superfield have the following
form:
\bea
&& D_\alpha (D\bar{D})W=-{i\over2}\partial_{\alpha\beta} D^\beta W~,
\lb{B8b}\\
&&(D\bar{D})^2 W= {1\over8}\partial^{\alpha\beta}(\partial_{\alpha\beta}
-iD_{\alpha\beta})W~.\lb{B8c}
\eea

The  components of the vector multiplet can be calculated as the 
$\theta{=}0$ parts of basic superfields and their spinor derivatives
\bea
&&\varphi(x)=W|_0=i\DbD V|_0~,\quad \lambda_\alpha(x)=
(D_\alpha W)|_0~,\lb{B9}\\
&&\bar{\lambda}_\alpha(x)=-\bar{D}_\alpha W|_0~,\qquad
A_{\alpha\beta}(x)=D_{\alpha\beta}V)|_0~,\lb{B10}       \\
&& F_{\alpha\beta}(x)=D_{\alpha\beta}W|_0~,\qquad G(x)=i(D\bar{D}) W|_0~,
\lb{B11}
\eea
where $A_{\alpha\beta}$ and $F_{\alpha\beta}$ are the $3D$-vector
field and its field-strength , $G$ is the real auxiliary component and
$\varphi, \lambda$ and $\bar{\lambda}$ are the physical scalar and spinor
fields. The scalar field appears as the $3D$ analog of the 3-rd component 
of the $4D$ gauge field.

The low-energy effective action of the $3D$ vector multiplet
describes a non-minimal interaction of the real scalar field with the
fermion and gauge fields. For the $U(1)$  gauge superfield 
$V$ this action has the following general form:
\be
S(W)=-{1\over2}\int d^7z H(W)~,\qquad \tau(W)=H^{\prime\prime}(W) > 0~,
\lb{B11b}
\ee
where $H(W) $ is the real convex function of $W$.  Note that the action 
conserves the $U_R(1)$ invariance.

The interesting feature of the $3D$ gauge theory is the existence
of the Chern-Simons term \cite{Si}
\be
S_{\s CS}= {ik\over4}\int d^7z V \DbD V~,\lb{B13}
\ee
where $k$ is some constant. The component form of this action contains
the the topological gauge term $\int d^3x A_\ab \partial^\alpha_\gamma
A^{\gamma\beta}$. Note that the non-abelian generalization of this term 
has been constructed in ref.\cite{ZP1}.

The $3D$ linear multiplet is dual to the chiral multiplet $\phi$. 
The Legendre transform describing  this duality is
\be
S[B,\Phi]=-{1\over2}\int d^7z [H(B)-\Phi B]~,\lb{B12}
\ee
where $B$ is the real unconstrained superfield and $\Phi{=}\phi +
\bar{\phi}$. Varying the Lagrange multipliers $\phi$ and $\bar{\phi}$ 
one can obtain the constraints \p{B8}.

Using the solution of the algebraic $B$-equation
\be
H^\prime(B)\equiv f(B)= \Phi \lb{B14}
\ee
one can pass to the self-interaction of the chiral superfields
\bea
&& B~\Rightarrow~B(\Phi)=f^{-1}(\Phi)~,\lb{B15}\\
&& S(\Phi)=-{1\over2}\int d^7z\hat{H}(\Phi)~,\lb{B16}\\
&& \hat{H}(\Phi)=H[B(\Phi)]-\Phi B(\Phi)~,\lb{B17}\\
&& {\partial\Phi\over\partial B}=\tau(B)~,\qquad {\partial^2\hat{H}\over
\partial\Phi^2}\equiv\hat{\tau}=-{1\over\tau}~.\lb{B17b}
\eea

The corresponding superfield equation of motion is
\be
\bDs \hat{H}^\prime(\Phi)=\hat{\tau}(\Phi)\bDs\bar{\phi}+{1\over2}
\hat{\tau}^\prime(\Phi)\bDa\bar{\phi}\bar{D}^\alpha\bar{\phi}=0~.
\lb{B18}
\ee

This chiral model describes the special case of the K\"{a}hler
supersymmetric $\sigma$-model which is completely determined by
the real function $H$ and possesses by construction the additional 
abelian isometry
\be
\phi~\rightarrow \phi+i\beta~,\lb{B19}
\ee
where $\beta$ is some real parameter.

\setcounter{equation}0
\section{ \lb{C}Difficulties with spontaneous breaking of 
supersymmetry }

Let us consider the spontaneous breaking of supersymmetry in the
non-minimal gauge model (\ref{B11b}) with the additional linear $FI$-term
\be
S_{\s FI}={1\over2} \xi\int d^7z V~,\lb{C0}
\ee
where $\xi$ is a constant of the dimension $-1$.  Varying the superfield
$V$ one can derive the corresponding superfield equation of motion
\be
-i(D\bar{D}) H^\prime(W) +\xi=-i\tau(W)(D\bar{D})W-{i\over2}\tau^\prime(W)
D^\alpha W\bar{D}_\alpha W+\xi=0~,\lb{C1}
\ee
where
\be
\tau(W)=H^{\prime\prime}(W)~,\qquad \tau^\prime(W)=H^{\prime\prime\prime}
(W)~.\lb{C1b}
\ee

The spinor derivatives of this superfield equation generate the
component equations of motion of different dimension
\bea
&&D_\alpha(D\bar{D})H^\prime=-{i\over2}\tau \pab D^\beta W +{1\over2}
\tau^\prime  D_\alpha W(D\bar{D})W \nn\\
&&+{1\over2}\tau^\prime D^\beta W (D_\ab
+{i\over2}\pab )W-{1\over4}\tau^{\prime\prime}\bar{D}_\alpha W D^\beta W
D_\beta W-{1\over2}\tau^\prime\bar{D}_\alpha W\Ds W =0~,\lb{C1c}
\eea
where the last term vanishes due to the constraint \p{B8}. The vacuum
solutions can be analysed with the help of the equation
\be
 (D\bar{D})^2H^\prime (W)=0~.
\lb{C1d}
\ee

We shall study the constant solutions of the equation of motion using
the following vacuum ansatz:
\be
V_0=2i\tbt a -2\ts\bts G~,\qquad W_0=a + 2i\tbt G~,\lb{C2}
\ee
where $a$ and $G$ are constants.
The lowest vacuum components of the equations (\ref{C1}) and (\ref{C1d})
 read
\bea
&&G\tau(a)-\xi=0~,                   \lb{C3}\\
&& G^2\tau^\prime(a)=0~.\lb{C3b}
\eea
The non-trivial solution $G_0{\neq}0$ is possible for the quadratic
function $H$ only.

It is useful to consider the real $3D$ spinors $\lambda^\alpha_i=
\overline{\lambda^\alpha_i}$
\be
\lambda^\alpha={1\over\sqrt{2}}(\lambda^\alpha_1 +i\lambda^\alpha_2)
 \lb{C4}
\ee
and the corresponding real spinor parameters of the $N{=}2$ supersymmetry
\be
\epsilon^\alpha={1\over\sqrt{2}}(\epsilon^\alpha_1 +i\epsilon^\alpha_2)~.
 \lb{C5}
\ee

It is clear that the constant solution (\ref{C2}) can only break 
spontaneously {\it both} supersymmetries
\be
\delta_\epsilon\lambda^\alpha=iG_0\bar{\epsilon}^\alpha~,\qquad
\delta_\epsilon\bar{\lambda}^\alpha=-iG_0\epsilon^\alpha \lb{C6}
\ee
since it generates two real Goldstone fermions.

Thus, the full spontaneous breakdown of the $N{=}2$ supersymmetry
is possible only for the free  theory with the $W^2(V)$-interaction
and the $FI$ term. The partial spontaneous breaking is forbidden  if one
 uses the vector multiplet satisfying the standard constraint (\ref{B8}).

Let us estimate the role of the Chern-Simons term \p{B13} in the 
vacuum equations.
Varying the action $S(W)+S_{\s CS}+S_{\s FI}$ one can obtain the modified
 equation of motion
\be
-i(D\bar{D})H^\prime +k W +\xi=0\lb{CSeq1}~.
\ee

This superfield equation produces the following modified vacuum
equations:
\bea
&&G\tau(a)-k a-\xi=0~,\lb{vac1}\\
&& G^2\tau^\prime(a) -kG=0~.\lb{vac2}
\eea

The scalar potential of this model is
\be
{\cal V}_k(a)={1\over2\tau(a)}(\xi+ka)^2~.\lb{sc1}
\ee

For the arbitrary  function $\infty >\tau(a)>0$ this potential has the
unique manifestly supersymmetric minimum $a=-\xi k^{-1}$. Thus, even the
free $SBGS$ solution $\tau=const$ disappears in the presence of the
$CS$-term.

\setcounter{equation}0
\section{\lb{D}Partial spontaneous breakdown of  supersymmetry}

We shall define the modified Goldstone-type constraints for the $3D$
vector multiplet by the analogy with  refs.\cite{IKP,IZ}
and show that the partial spontaneous breaking of the $D{=}3,~N{=}2$
supersymmetry  is possible for the non-trivial gauge
interaction in the framework of this approach.

Consider the following deformation of the constraints (\ref{B8}):
\be
(D)^2\hat{W}=C~,\qquad (\bar{D})^2\hat{W}=\bar{C}~,\lb{D1}
\ee
where $C$ and $\bar{C}$ are some constants. These relations break
manifestly the $U_R(1)$ invariance.

The solution of these constraints can be constructed by analogy with
Eq.(\ref{B7})
\be
\hat{W}=i\DbD V+(\theta)^2C
+(\bar{\theta})^2\bar{C}~.\lb{D2}
\ee
This $LGM$ superfield contains new constant auxiliary
structures which change radically the matrix of the vacuum fermion
transformations
\bea
&&\delta_\epsilon\lambda^\alpha=-C\epsilon^\alpha+iG_0
\bar{\epsilon}^\alpha~,
\lb{D3}\\
&&\delta_\epsilon\bar{\lambda}^\alpha=
-iG_0\epsilon^\alpha-\bar{C}\bar{\epsilon}^\alpha~.\lb{D4}
\eea

It is evident that the $PSBGS$-condition corresponds to the degeneracy
of these transformations
\be
C\bar{C}-G^2_0=0~.\lb{D5}
\ee
In this case one can choose the single real Goldstone spinor field as
some linear combination of $\lambda^\alpha_i$. For the case of the
pure imaginary constant $C$ the $LG$ fermion can be identified
with $\lambda^\alpha_2$.

It should be stressed that the shifted  quantity
$W(V)=i\DbD V$ in Eq.(\ref{D2}) is not a standard superfield
\be
\delta_\epsilon\hat{W}=i\epsilon^\alpha_k Q_\alpha^k \hat{W}~
\rightarrow~\delta_\epsilon W(V)=
C\epsilon^\alpha\theta_\alpha -\bar{C}\bar{\epsilon}^\alpha
\bar{\theta}_\alpha+i\epsilon^\alpha_k Q_\alpha^k W(V)~,\lb{D6}
\ee

The algebra of these transformations is not changed on the gauge-invariant 
superfield
\be
[\delta_\eta ,\delta_\epsilon] W(V)=\epsilon^\alpha_k\eta^\beta_l
\{Q_\alpha^k,Q_\beta^l\}W(V)~.\lb{D10}
\ee
The transformation of the $LGM$-prepotential $V$ will be considered
in the end of this section.

The action of the $LGM$-superfield (\ref{D2}) has the following form:
\be
\hat{S}(V)=-{1\over2}\int d^7z [H(\hat{W})-\xi V]\lb{D11}
\ee
and depends on three constants $\xi, C$ and $\bar{C}$. 

The non-derivative terms in the component Lagrangian
\be
{1\over2}(G^2-|C|^2)\tau(\varphi)-\xi G~,\lb{Lagr}
\ee
produce the following scalar potential
\be
{\cal V}(\varphi)={1\over2}[|C|^2\tau(\varphi)+\xi^2\tau^{-1}(\varphi)]~.
\lb{D15}
\ee

The vacuum equations of this model are
\bea
&&G\tau(a)-\xi=0~,    \lb{D12}\\
&& (G^2-|C|^2)\tau^\prime(a)=0~.\lb{D13}
\eea

The $PSBGS$ solution (\ref{D5}) arises for the non-trivial interaction
$\tau^\prime(a){\neq}0$. This solution determines the minimum point 
$a_{\s0}$ of this model
\be
\tau(a_{\s0})=\frac{|\xi|}{|C|}~.\lb{D14}
\ee

The vacuum auxiliary field can be calculated in the point $a_{\s0}$
\be
G_{\s0}=\frac{\xi}{\tau(a_{\s0})}=\pm|C|~.\lb{D16}
\ee
Using the $U_R(1)$ transformation one can choose the pure imaginary
constant $C\rightarrow c=i|c|$ (without the loss of generality) then
\be
G_{\s0}=-ic=|c|~.\lb{D17}
\ee

This choice corresponds to the following decomposition of the
$LGM$-superfield (\ref{D2})
\bea
&&\hat{W}=W_s(V_s) +2i|c|\ttt~,\lb{D18}\\
&& W_s(V_s)={i\over4}(D^{1\alpha}D^1_\alpha+D^{2\alpha}D^2_\alpha)
V_s~.\lb{D18b}
\eea
where $V_s$ is the shifted $LGM$-prepotential which has  the vanishing
vacuum solution for the auxiliary component. It is evident that this 
representation breaks spontaneously the 2-nd supersymmetry only.

Let us consider now the  supersymmetry transformations of $W_s$ and
$V_s$
\bea
&&\delta_\epsilon W_s=i\epsilon^\alpha_kQ_\alpha^k \hat{W}=
-2i|c|\epsilon^\alpha_2\theta_{2\alpha}+ i\epsilon^\alpha_k 
Q_\alpha^k W_s~,
\lb{D19}\\
&& \delta_\epsilon V_s=\Delta(\epsilon,\theta)+
i\epsilon^\alpha_k Q_\alpha^k V_s
~,\lb{D20}\\
&& \Delta(\epsilon,\theta)=2|c|\epsilon^\alpha_2
\theta_{2\alpha}(\theta^\beta_1\theta_{1\beta})=-2\sqrt{2}|c|i
\epsilon^\alpha_2[\bar{\theta}_\alpha\ts +\theta_\alpha\bts]~.\lb{D20b}
\eea

The supersymmetry algebra of the $V_s$-transformations is essentially
modified by the analogy with the transformations of the prepotentials
in refs.\cite{FGP,IZ}
\bea
&&[\delta_\eta ,\delta_\epsilon] V_s\equiv
\epsilon^\alpha_k\eta^\beta_l\{\tilde{Q}_\alpha^k,\tilde{Q}_\beta^l\}V_s
\nn\\
&& =4|c|(\epsilon^\alpha_2\eta^\beta_1-\eta^\alpha_2\epsilon^\beta_1)
\theta_{1\beta}\theta_{2\alpha}+
\epsilon^\alpha_k\eta^\beta_l\{Q_\alpha^k,Q_\beta^l\}V_s~,\lb{D21}
\eea
where $\tilde{Q}_\alpha^k$ are the generators of the modified
transformations.

The modified part of the supersymmetry transformation has the following
form:
\be
\{\tilde{Q}_\alpha^1,\tilde{Q}_\beta^2\}_{mod}V_s=4|c| \theta_{1\alpha}
\theta_{2\beta}=
4i|c|\Theta_\ab+2i|c|\varepsilon_\ab [\ts+\bts]~.
\lb{D22}
\ee
It should be stressed that both terms in this anticommutator can be
decomposed as a sum of chiral and anti-chiral functions
 and do not contribute to the Lie bracket on the superfield $W_s$
 \be
\Theta^\ab=-{i\over2}(x^\ab_L-x^\ab_R)~,\qquad x^\ab_R\equiv
(x^\ab_L)^\dagger~.\lb{D23b}
\ee

The modified anticommutator contains  the
additional vector and scalar generators $T_\ab,T $ and $\bar{T}$
\bea
&&\{\tilde{Q}_\alpha^1,\tilde{Q}_\beta^2\}=\varepsilon_\ab(T+\bar{T})
+T_\ab~.\lb{D23}\\
&&  T_\ab V_s=4i|c|\Theta_\ab~,\quad TV_s=2i|c|\ts~.
\lb{D24}
\eea
The additional generators belong to the infinite Lie algebra of the
$U(1)$-gauge transformations which arises in the $(x,\theta)$-decomposition
of the chiral gauge parameters $\Lambda$ \p{B5}. These generators vanish
on the gauge invariant quantity $W_s$. One should also include  in the
modified $N{=}2$ supersymmetry algebra all nontrivial commutators of the
$T$ generators with the spinor generators $\tilde{Q}_\alpha^k$.

Consider the spinor gauge connection
\be
A_\alpha(z)=\Da V_s~,\qquad \delta_\Lambda A_\alpha=\Da \Lambda
\lb{D25}
\ee
in the chiral representation $(\bar{A}_\alpha{=}0)$.
The inhomogeneous term in the modified supersymmetry transformation of 
this gauge superfield has the following form:
\be
\delta A_\alpha=-2\sqrt{2}i|c|[\epsilon_{2\alpha}\bts-
\theta_\alpha\epsilon^\beta_2\bar{\theta}_\beta]+
i\epsilon^\alpha_k Q_\alpha^k A_\alpha~. \lb{D26}
\ee

It should be remarked that the minimal interaction of the charged chiral
superfields with the $LGM$-prepotential $V_s$ breaks the supersymmetry.
The analogous problem of the $LGM$ interaction with the charged matter
appears also in the $PSBGS$ model with $D{=}4,~N{=}2$ supersymmetry
\cite{IZ}.

\setcounter{equation}0
\section{\lb{E}The $3D$ chiral interaction with the partial 
 breaking}

 The general effective action
of the chiral superfields $\phi_i$ ($i$ is some internal index)
can be written as follows:
\be
\int d^4x d^4\theta K(\phi_k,\bar{\phi}_k) +[\int d^4x d^2\theta
P(\phi_i)+\mbox{c.c.}]~,\lb{E0}
\ee
where $K$ is the K\"{a}hler potential and $P$ is the chiral
superfield potential.

The existence of the non-trivial $SBGS$ solution implies the degeneracy
of the matrix $\partial_i\partial_k P$.
The vacuum equation for the single chiral superfield $\phi$ may have
the non-vanishing  $SBGS$ solution only in the trivial case of the linear
function $P(\phi)$ and the free K\"{a}hler potential $K=\phi\bar{\phi}$.

We shall show that the spontaneous breaking of supersymmetry is possible
for the non-trivial interaction of the $LG$ chiral superfield 
which possesses the inhomogeneous supersymmetry transformation.
Let us consider the dual picture for the $PSBGS$ gauge model with the
$FI$-term (\ref{D11})
\be
\hat{S}(B,\phi,\bar{\phi})=-{1\over2}\int d^7z [H(B) - B\hat{\Phi}]
-{1\over2}[\bar{C}\int d^5\zeta\phi +\mbox{c.c.}]~,
\lb{E1}
\ee
where the modified constrained $LG$ superfield is introduced
\be
\hat{\Phi}\equiv \phi+\bar{\phi}+ 2i\xi\tbt~,\qquad \DbD\hat{\Phi}=-i\xi~.
\lb{E1b}
\ee

Varying the chiral and antichiral Lagrange multipliers $\phi$ and
$\bar{\phi}$ one can obtain the $LGM$-constraints (\ref{D2}) on the 
superfield $B$ and then pass to the gauge phase $B\rightarrow \hat{W}(V)$ 
where the $\tbt$-term in $\hat{\Phi}$ transforms to the $FI$-term.

The algebraic $B$-equation
\be
H^\prime(B)\equiv f(B)=\hat{\Phi}
~,\lb{E2}
\ee
provides the transform to the `chiral' phase
\be
B~\rightarrow~f^{-1}(\hat{\Phi})\equiv\hat{B}(\hat{\Phi})~.\lb{E3}
\ee

The transformed chiral action is
\bea
&&\hat{S}(\hat{\Phi})=-{1\over2}\int d^7z  \{\hat{H}(\hat{\Phi})
+[\bar{C}\ts +\mbox{c.c.}]\hat{\Phi}\}~,\lb{E4}\\
&& \hat{H}(\hat{\Phi})=H[\hat{B}(\hat{\Phi})]-\hat{\Phi}\hat{B}
(\hat{\Phi})\lb{E5}
\eea
 The linear terms with $C$ and $\bar{C}$ break
the  $U_R(1)$-symmetry \p{A4c}, however, this action is
invariant with respect to the isometry transformation \p{B19}.

It should be underlined that the $LG$-superfield $\hat{\Phi}$ 
transforms homogeneously, while the supersymmetry transformation of the 
$LG$-chiral Lagrange multiplier $\phi$ contains the inhomogeneous term
\be
\delta_\epsilon\phi=
-i\xi(\theta^\alpha\bar{\epsilon}_\alpha)+i\epsilon^\alpha_k Q_\alpha^k
\phi~.
\lb{E6}
\ee

 The action $\hat{S}$ is invariant with respect to the $LG$ representation
of the $N{=}2$ supersymmetry, since the 1-st term of this action depends 
manifestly on  the covariant superfield $\hat{\Phi},$ and the 2-nd one is
invariant due to the linear $\theta$-dependence of the inhomogeneous part 
of $\delta_\epsilon\phi$.

Consider the $\theta$-decomposition of the $LG$-chiral superfield
\be
\phi=A(x_L) +\theta^\alpha \psi_\alpha(x_L)+(\theta)^2
F(x_L)~,\lb{E7}
\ee
where  $x_L$ is the shifted coordinate of the chiral basis.

The Lie bracket of the modified supersymmetry transformation \p{E6}
\be
[\delta_\eta,\delta_\epsilon]\phi=i\xi
(\epsilon^\alpha\bar{\eta}_\alpha -\eta^\alpha\bar{\epsilon}_\alpha)+
\epsilon^\alpha_k\eta^\beta_l\{Q_\alpha^k,Q_\beta^l\}\phi
\lb{E8}
\ee
 contains the composite central charge parameter corresponding to the
the following action of the generator $Z$ on the chiral superfield:
\be
Z \phi=\xi~,\qquad ( Z\bar{\phi} =-\xi )
~.\lb{E8b}
\ee
Thus, the Goldstone boson field $\mbox{Im}\,A(x)$ for the central-charge 
transformation appears in this model.
It should be remarked that the isometry transformation \p{B19} in the 
chiral model without $PSBGS$ cannot be identified with the central charge.

It is interesting that we can define the deformed chiral superfield
\bea
&&\phi_\xi=\phi +i\xi\tbt=e^{i\tbt Z}\phi~,\qquad Z^2 \phi =0~,\nn\\
&& \bcDa \phi_\xi =( \bDa -{i\over2}\theta_\alpha Z)\phi_\xi=0\lb{defch1}
\eea
satisfying the unusual covariant condition. 

The superfield equation of motion for the action (\ref{E4}) 
\be
 (\bar{D})^2\hat{H}^\prime(\hat{\Phi})+\bar{C}=0\lb{E12}
\ee
generates the vacuum component  equations
\bea
&& \bar{F}\hat{\tau}(b) + \bar{C}=0~,\qquad b=A+\bar{A}\lb{E13}\\
&& (|F|^2-\xi^2)\hat{\tau}^\prime(b)=0~,\lb{E14}\\
&&  \hat{\tau}=\hat{H}^{\prime\prime}=-\tau^{-1}~.\lb{E15}
\eea

The scalar potential of this model depends on the one real scalar
component only
\be
{\cal V}(b)={1\over2}[\xi^2\hat{\tau}(b)+|C|^2\hat{\tau}^{-1}(b)]~.
\lb{E16}
\ee
The minimum point $b_{\s0}$ of this potential can be defined by the
equation
\bea
&&{\cal V}^\prime={1\over2}\hat{\tau}^\prime(b)[\xi^2 - |C|^2
\hat{\tau}^{-2}(b)]=0~,\lb{E17}\\
&& \tau^2(b_{\s0})=\hat{\tau}^{-2}=\xi^2|C|^2~.\lb{E18}
\eea
using the condition $\tau^\prime(b){\neq}0$.

The vacuum transformations of the spinor components of the $LG$
superfields $\phi$ and $\bar{\phi}$ have the following form:
\bea
&&\delta_\epsilon\psi^\alpha=F_0\epsilon^\alpha-
i\xi\bar{\epsilon}^\alpha~,
\lb{E9}\\
&&\delta_\epsilon\bar{\psi}^\alpha=i\xi\epsilon^\alpha
+\bar{F}_0\bar{\epsilon}^\alpha~.\lb{E10}
\eea

The vacuum solution $|F_0|^2=\xi^2$ corresponds to the degeneracy 
condition for these transformations. The choice $F_0=i\xi$ breaks the 
2-nd supersymmetry.

Thus, the non-trivial interaction of the $LG$-chiral superfield
$\phi$ provides the partial spontaneous breaking of the $D{=}3,~N{=}2$
supersymmetry. This phenomenon has been analysed also in the formalism
of the $D{=}3,~N{=}1$ Goldstone-type superfields \cite{IK}.

\setcounter{equation}0
\section{\lb{F}Passing to $N{=}1$ superfields}

Let us assume that the spinor coordinates $\tao$ parameterize $N{=}1$
superspace, and the generators $Q_\alpha^1$ form the corresponding
subalgebra of the $N{=}2$ supersymmetry.
The complex chiral coordinates $\zeta$ \p{A9b} can be written via the
real spinor coordinates
\bea
&&x_L^\ab=x^\ab +{1\over2}(\tao\theta^\beta_2 +\alpha\leftrightarrow
\beta)~,\lb{F1}\\
&&\ta={1\over\sqrt{2}}(\tao +i\tat )~.\lb{F1b}
\eea

We shall use the relations
\bea
&&\Ds={1\over2}\Doo -{1\over2}\Dtt -i(D^1D^2)~,\qquad (D^iD^k)\equiv
{1\over2}D^{i\alpha}D^k_\alpha~,\lb{F2}\\
&& \Dao\Dbo={i\over2}\pab+\varepsilon_\ab\Doo~,\qquad \{\Dao,\Doo\}=0~,
\lb{F3}\\
&&[\Dao,\Doo]=-i\pab D^{1\beta}~,\qquad  \Doo^2={1\over8}\Box_3
~,\qquad\Box_3=\partial^\ab\pab~.\lb{F3b}
\eea

The chirality condition  in the real basis
\be
\bDa \phi=(\Dao+i\Dat )\phi=0
\lb{F4}
\ee
 can be solved via the complex unrestricted $N{=}1$ superfield 
$\chi$
\be
\phi=\chi(x,\theta_1) +i\tat\Dao \chi(x,\theta_1) +\ttt \Doo
\chi(x,\theta_1)~.
\lb{F5}
\ee
To prove the chirality in the $N{=}1$ representation one should
 use  Eqs.(\ref{F3},\ref{F3b}) and the relation
\be
\Dat \chi(x,\theta_1)={i\over2}\theta^\beta_2\pab \chi(x,\theta_1)~.
\lb{F6}
\ee

Using Eq.\p{F2} one can readily obtain the relation between the chiral
and $N{=}1$ integrals
\be
\int d^3x\Ds \phi =\int d^3x \Doo \chi(x,\theta_1)~, 
\lb{F6b}
\ee
where $d^2\theta_1{=}\Doo$ is the imaginary spinor measure of the $N{=}1$
superspace.

The transformation \p{E6} has the following $N{=}1$ decomposition:
\be
\delta\phi=-{1\over2}\xi\tao(\epsilon_{2\alpha}+i\epsilon_{1\alpha})
+{1\over2}\xi\tat(\epsilon_{1\alpha}-i\epsilon_{2\alpha})+
\epsilon^\alpha_2
(-\partial^2_\alpha+{i\over2}\theta^\beta_2\pab)\phi +
i\epsilon^\alpha_1Q^1_\alpha\phi \lb{F7}
\ee
and generates the corresponding transformation of the complex $N{=}1$
superfield:
\be
\delta\chi=-{1\over2}\xi\tao(\epsilon_{2\alpha}+i\epsilon_{1\alpha})
-i\epsilon^\alpha_2\Dao \chi+
i\epsilon^\alpha_1Q^1_\alpha \chi~.\lb{F8}
\ee

Consider the $\theta_2$-decomposition of the basic superfield \p{E1b}
of the chiral $PSBGS$ model
\bea
&&\hat{\Phi} =\chi+\bar{\chi}+i\tat\Dao(\chi-\bar{\chi})+
\ttt\Doo(\chi+\bar{\chi})+i\xi[\too+\ttt]\nn\\
&&=\Sigma+\tat\Dao \rho+
\ttt[\Doo\Sigma+2i\xi]~,\lb{F10}\\
&&\Sigma(x,\theta_1)=\chi+\bar{\chi}+i\xi\too~,\qquad 
\rho=i\chi-i\bar{\chi}\lb{F11}
\eea
where $\Sigma$ is the standard real $N{=}1$ superfield and $\rho$ is
the real Goldstone superfield for the 2-nd supersymmetry
\bea
&&\delta\Sigma= -\epsilon^\alpha_2\Dao \rho+
i\epsilon^\alpha_1Q^1_\alpha \Sigma~.\lb{F12}\\
&&\delta\rho=-i\xi\epsilon_2^\alpha\theta_{1\alpha}
+\epsilon^\alpha_2\Dao \Sigma+
i\epsilon^\alpha_1Q^1_\alpha \rho~.\lb{F13}
\eea

The analogous transformations of $N{=}1$ superfields have been proposed
in ref.\cite{IK}. The authors of this work have shown that the additional
superfield  can be constructed in terms of the spinor derivative
of the Goldstone superfield $\rho$ in order to built the supermembrane
action. The massive degrees of freedom in our approach can be removed 
using the covariant condition
\be
\hat{\Phi}^2=0~,\lb{F16b}
\ee
which allows us to construct $\Sigma$ via $\Dao \rho$ by analogy with
the similar construction in the $D{=}4,~N{=}2$ theory \cite{RT}.

The superfield $\rho$ possesses also the central-charge transformation
induced by the corresponding transformation of the chiral superfield
\p{E8b}.
%\be
%\delta_Z\rho(x,\theta_1)=-\xi\beta~.\lb{F14}
%\ee

 Our $N{=}2$ action \p{E4} can be rewritten via the both 
$N{=}1$ components of $\hat{\Phi}$ 
\bea
&&-{1\over2}\bar{C}\int d^3x\Ds\phi +\mbox{c.c.}={1\over2}\int d^3x 
d^2\theta_1[(C-\bar{C}) \Sigma+i(C+\bar{C})\rho]+\mbox{const}~,
\lb{F15}\\
&&-{1\over2}\int d^7z \hat{H}(\hat{\Phi})=-{1\over2}\int d^3x \Doo\Dtt
 \hat{H}(\hat{\Phi})\nn\\
&&={1\over2}\int d^3x d^2
\theta_1 \{[2i\xi+\Doo\Sigma] \hat{H}^\prime
(\Sigma)+{1\over2}\hat{\tau}(\Sigma)D^{1\alpha}\rho\Dao\rho\}~,\lb{F16}
\eea
Note that these  integrals, including the linear in $\rho$ term, are 
invariant with respect to the  $N{=}2$ supersymmetry transformations 
(\ref{F12},\ref{F13}).

Let us analyse the $N{=}1$ decomposition of the gauge prepotential
\be
V_s(x,\theta_1,\theta_2)=\kappa(x,\theta_1)+i\tat V_\alpha(x,\theta_1)+
i\ttt M(x,\theta_1)\lb{F17}
\ee
and the chiral gauge parameter
\be
\Lambda =[1+i\tat\Dao +\ttt\Doo]\lambda(x,\theta_1)~.\lb{F18}
\ee

The gauge transformations of the $N{=}1$ components are
\bea
&& \delta_\lambda \kappa=\lambda+\bar{\lambda}~,\lb{F19}\\
&& \delta_\lambda V_\alpha=\Dao(\lambda-\bar{\lambda})~,\lb{F20}\\
&&\delta_\lambda M=-i\Doo(\lambda+\bar{\lambda})~.\lb{F21}
\eea
Thus, $\kappa$ is a pure gauge degree of freedom, $V_\alpha$ is the
$N{=}1$ gauge superfield, and $M$ is the scalar $N{=}1$  component of
the $N{=}2$ supermultiplet.

The 2-nd supersymmetry transformations of the $N{=}1$ superfields have
the following form
\bea
&& \delta_2\kappa=-i\epsilon^\alpha_2 V_\alpha~,\lb{F22}\\
&& \delta_2 V_\alpha =-\epsilon_{2\alpha}[M+4i|c|\too]
 -{1\over2}\epsilon^\beta_2\pab\kappa~,\lb{F23}\\
&&\tilde{\delta}_2 M={1\over2}\epsilon^\alpha_2\pab V^\beta~.\lb{F24}
\eea
The deformation of the supersymmetry algebra \p{D22} can be studied
also in this representation.

Consider the $N{=}1$ decomposition of the linear superfield \p{D18b}
\bea
&&W_s(V_s)={i\over2}[\Doo+\Dtt]V_s
=w +i\tat F_\alpha(V)-\ttt\Doo w~,\lb{F25}\\
&& [\Doo-\Dtt]W_s=0~,\qquad (D^1D^2)W_s=0~.\lb{F25b}
\eea
where  the gauge-invariant scalar and spinor superfields are defined
\bea
&&w={1\over2}[M+i\Doo\kappa]~,\lb{F26}\\
&& F_\alpha(V) ={i\over2}\Doo V_\alpha+{1\over4}\pab V^\beta~,\qquad
D^\alpha F_\alpha =0~. \lb{F27}
\eea

The Goldstone transformation of $W_s$ \p{D19} produces the following
$\epsilon_2$-transformations of the $N{=}1$ superfields:
\bea
&&\delta w=-i\epsilon_2^\alpha F_\alpha~,\lb{F28}\\ 
&&\delta F_\alpha =-\epsilon_{2\alpha}[2|c|+i\Doo w]
+{i\over2}\epsilon^\alpha_2\pab w~.\lb{F29}
\eea

The spinor superfield strength $F_\alpha$ is analogous to the Goldstone
spinor superfield of ref.\cite{IK}. It describes the Goldstone
degree of freedom of the $D2$-brane, and the superfield $w$
corresponds to the massive degrees of freedom. Our construction
introduces the $N{=}1$ gauge superfield $V_\alpha$ as the basic
object of this model and allows us to study the modification
of the supersymmetry algebra on the gauge fields of the $D2$-brane. It
 is not difficult to rewrite the $N{=}2$ action \p{D11} in terms of the 
$N{=}1$ superfields.

\vspace{0.5cm}

\noindent{\large\bf  Acknowledgment}\\

The author is grateful to  E.A. Ivanov, S.O. Krivonos, A.A. Kapustnikov,
 O. Lechtenfeld and, especially, J. Lukierski for stimulating discussions 
and thanks J. Lukierski and Z. Popowicz for the kind hospitality
in Wroclaw University where the main part of this work has been made.
The work  is partially supported by the grants RFBR-99-02-18417, 
INTAS-93-127-ext, INTAS-96-0308, by the Bogoliubov-Infeld programme  and by 
the Uzbek Foundation of Basic Research, contract N 11/27.

\end{document}